\documentclass[twocolumn,showpacs,preprintnumbers,amsmath,amssymb]{revtex4}
\usepackage{graphicx}% Include figure files
\usepackage{dcolumn}% Align table columns on decimal point
\usepackage{bm}% bold math
\begin{document}
\title{Strong enhancement of the valley splitting in a 2D electron
system in silicon}
\author{V.~S. Khrapai, A.~A. Shashkin, and V.~T. Dolgopolov}
\affiliation{Institute of Solid State Physics, Chernogolovka, Moscow
District 142432, Russia}
\begin{abstract}
Using magnetocapacitance data, we directly determine the chemical
potential jump in a strongly correlated 2D electron system in silicon
when the filling factor traverses the valley gap at $\nu=1$ and
$\nu=3$. The data yield a valley gap that is strongly enhanced
compared to the single-particle value and increases {\em linearly}
with magnetic field. This result has not been explained by existing
theories.
\end{abstract}
\maketitle

A two-dimensional (2D) electron gas in (100)-silicon
metal-oxide-semiconductor field-effect transistors (MOSFETs) is a
unique double-layer electron system with strong interlayer
interactions. Indeed, the valley index is identical with the isospin
quantum number, and one can therefore try to apply to this system a
good deal of recent theoretical work on double-layer electron
systems. In a double-layer system with small layer separation
compared to the interelectron distance in either layer, strong
interlayer correlations are predicted to give rise to the appearance
of both novel ground states and novel excitations \cite{sondhi,yang}.
In quantizing magnetic fields strong enough to fully polarize real
electron spins, two cases as determined by the isospin system
symmetry are discussed: (i) at finite layer separation the symmetry
is U(1) corresponding to an ``easy-plane'' anisotropy and (ii) at
vanishing layer separation the symmetry transforms to SU(2) and the
anisotropy disappears. In the first (second) case the lowest energy
charge-carrying excitations are merons (skyrmions) that are isospin
textures with the charge $e/2$ ($e$), where $e$ is the electron
charge \cite{sondhi,yang}. There is little doubt that the valley
splitting in the 2D electron system in silicon MOSFETs should be of
many-body origin at least for lowest odd filling factors because the
electron-electron interactions (and correlations) in this system are
strong; particularly, in accessible magnetic fields, the Coulomb
energy exceeds significantly the cyclotron energy. However, this is
the strongly interacting limit in which existing theories are not
valid and, therefore, they cannot be directly applied to silicon
MOSFETs. The origin of the excitations for the valley splitting is
unknown so far.

Experimental investigations of the valley splitting were performed
largely \cite{qq} at high filling factor $\nu\ge 9$ based on analysis
of the beating pattern of Shubnikov-de Haas oscillations in tilted
magnetic fields \cite{koehler,nicholas,wakabayashi,pud}. The gap
value, its linear dependence on substrate bias \cite{nicholas}, and
its insensitivity to parallel magnetic field \cite{nicholas1} are
consistent with single-particle theoretical considerations for an
asymmetric potential well that contains a 2D electron gas 
\cite{afs,rem}.

In this paper, we perform, for the first time, low-temperature
measurements of the chemical potential jump across the valley gap at
the lowest filling factors $\nu=1$ and $\nu=3$ in a 2D electron
system in silicon using a magnetocapacitance technique. The valley
splitting is found to exceed strongly the single-particle value,
decaying with filling factor. Unexpectedly, the data are best
described by a {\em linear} increase of the valley gap with magnetic
field, which is similar to the proportional magnetic field dependence
of the enhanced spin gap for the 2D electrons in AlGaAs/GaAs
heterostructures \cite{usher,aristov}.

Measurements were made in an Oxford dilution refrigerator with a base
temperature of $\approx 30$~mK on high-mobility (100)-silicon MOSFETs
(with a peak mobility close to 2~m$^2$/Vs at 4.2~K) having the
Corbino geometry with diameters 250 and 660~$\mu$m. The gate voltage
was modulated with a small ac voltage 15~mV at frequencies in the
range 2.5 -- 25~Hz and the imaginary current component was measured
with high precision using a current-voltage converter and a lock-in
amplifier. Care was taken to reach the low frequency limit where the
magnetocapacitance is not distorted by lateral transport effects. A
dip in the magnetocapacitance at integer filling factor is directly
related to a jump of the chemical potential across a corresponding
gap in the spectrum of the 2D electron system \cite{smith,pud86}.

\begin{figure}\vspace{1mm}
\scalebox{0.55}{\includegraphics{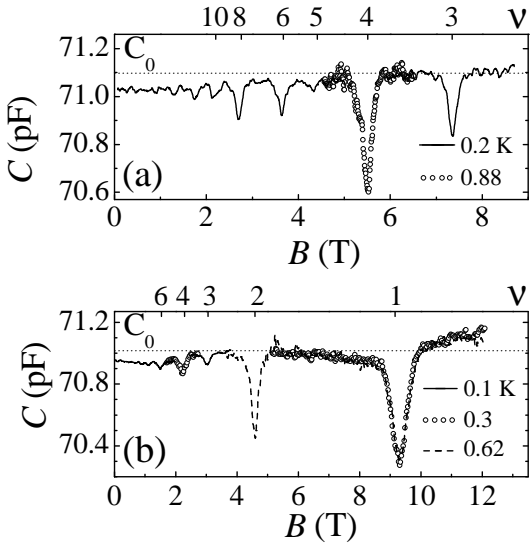}}
\caption{\label{fig1} Magnetocapacitance in the low frequency limit 
for electron densities $5.3\times 10^{11}$~cm$^{-2}$ (a) and 
$2.2\times 10^{11}$~cm$^{-2}$ (b). The level of the geometric 
capacitance $C_0$ is indicated by a dotted line.}
\end{figure}

Typical traces of the magnetocapacitance, $C(B)$, at different
electron densities, $n_s$, are displayed in Fig.~\ref{fig1}. It
oscillates as a function of filling factor, $\nu=hcn_s/eB$,
reflecting the modulation of the thermodynamic density of states
(DOS) in quantizing magnetic fields. Narrow dips in the
magnetocapacitance at integer $\nu$ are separated by broad maxima
where the value $C$ approaches in the high field limit the geometric
capacitance, $C_0$, between the gate and the 2D electrons, in
agreement with previous studies \cite{smith}:

\begin{equation}
\frac{1}{C}=\frac{1}{C_0}+\frac{1}{Ae^2D}, \label{C}\end{equation}
where $A$ is the sample area, $D=dn_s/d\mu$ is the thermodynamic DOS,
and $\mu$ is the chemical potential. We have checked that in the
range of electron densities used, the value $D(B=0)$, which is
calculated from the magnetocapacitance in the high field limit
($C_0$) and that in $B=0$, corresponds to $2m/\pi\hbar^2$, where
$m=0.19m_e$ and $m_e$ is the free electron mass. Note that at the
high field edge of the dip in $C(B)$, the capacitance $C$ can
overshoot $C_0$ forming a local maximum (so-called negative
compressibility effect \cite{pud89}), the effect being not so
pronounced at $\nu=3$ as at $\nu=1$, see Fig.~\ref{fig1}.

Integrating Eq.~(\ref{C}) over a dip at integer $\nu=\nu_0$ yields a
jump, $\Delta$, of the chemical potential between two neighboring
quantum levels \cite{pud86}

\begin{equation}
\Delta=\frac{Ae^3\nu_0}{hcC_0}\int_{\text{dip}}\frac{C_0-C}{C}dB,
\label{Delta}\end{equation}
where the integration over $B$ is equivalent to the one over $n_s$ as
long as the dip is narrow. We note that, experimentally, it is easier
to analyze $C(B)$ traces: while being independent of magnetic field,
the geometric capacitance $C_0$ increases slightly with $n_s$ as the
2D electrons are forced closer to the interface.

\begin{figure}\vspace{1mm}
\scalebox{0.55}{\includegraphics{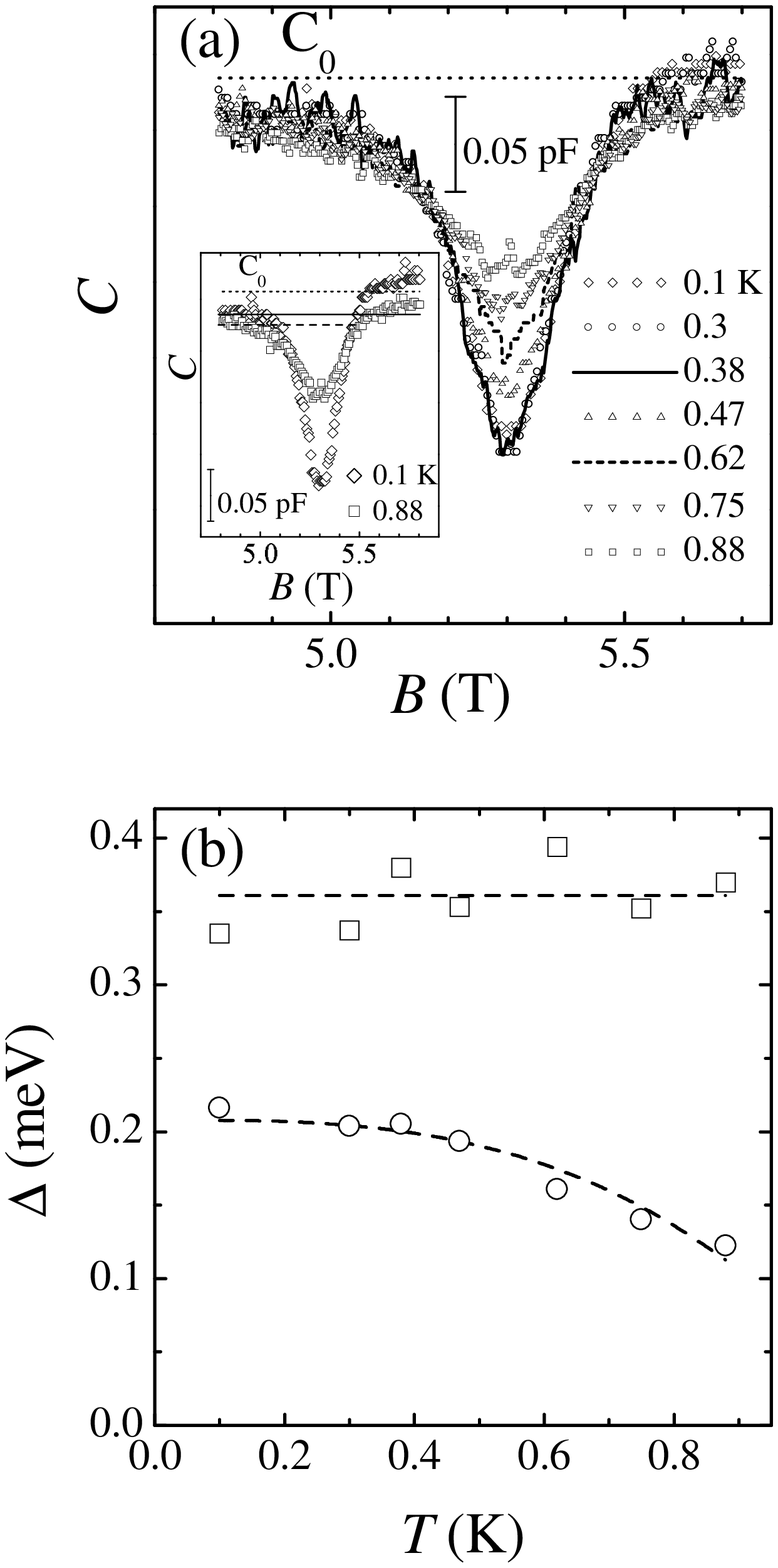}}
\caption{\label{fig2} The temperature dependence of the $\nu=3$ 
minimum in the magnetocapacitance at $n_s=3.85\times 
10^{11}$~cm$^{-2}$ (a) and of the chemical potential jump with 
(squares) and without (circles) the level width contribution (b). The 
dashed lines in (b) are guides to the eye. The inset shows the same 
minimum in $C(B)$ along with the shifted reference level (solid and 
dashed lines) used to determine the jump $\Delta$ of 
Eq.~(\ref{Delta}) when the geometric capacitance $C_0$ (dotted line) 
is not attained.}
\end{figure}

Apparently, if the capacitance $C$ does not reach $C_0$ (i.e., the
maximum thermodynamic DOS is insufficiently large), the jump $\Delta$
will be smaller than the level splitting by the level width. While
the jump $\Delta$ is still determined by the area of the dip in
$C(B)$ (see the inset of Fig.~\ref{fig2}(a)), we estimate the level
splitting including the level width contribution with the help of a
similar integration of Eq.~(\ref{C}) between the magnetic fields
$B_1=hcn_s/e(\nu_0+1/2)$ and $B_2=hcn_s/e(\nu_0-1/2)$.

In Fig.~\ref{fig2}(a), we show the $\nu=3$ minimum in the $C(B)$
curve for $n_s=3.85\times 10^{11}$~cm$^{-2}$ at different
temperatures. As the temperature, $T$, is lowered, the minimum
deepens appreciably and becomes narrower until it saturates in the
low $T$ limit, which is consistent with temperature smearing of the
thermodynamic DOS. In a similar way, the maximum capacitance
decreases with temperature departing from the geometric capacitance
$C_0$. In insufficiently high magnetic fields, the (positive)
difference between $C_0$ and $C$ at the low field edge of the dip is
larger compared to that at the high field edge. Therefore, we
determine the jump $\Delta$ by replacing the reference level $C_0$ in
Eq.~(\ref{Delta}) by the capacitance value at the low field edge of
the dip as depicted in the inset to Fig.~\ref{fig2}(a). This gives a
little underestimated values of $\Delta$ for such minima whose weakly
asymmetric shape persists more or less, irrespective of magnetic
field.

The chemical potential jump $\Delta(T)$ across the $\nu=3$ valley gap
for the data of Fig.~\ref{fig2}(a) with and without allowing for the
temperature-dependent level width is displayed in Fig.~\ref{fig2}(b).
The value $\Delta$ increases with decreasing $T$ and saturates in the
low temperature limit, this temperature dependence being weaker
compared to that of the minimum magnetocapacitance (and the minimum
$D$). The estimated valley splitting including the level width
contribution is practically independent of temperature, as expected.
It is clear that the experimental uncertainty of determination of the
valley splitting is the smallest in the low temperature limit where
the modulation of the thermodynamic DOS is maximal.

\begin{figure}\vspace{1mm}
\scalebox{0.55}{\includegraphics{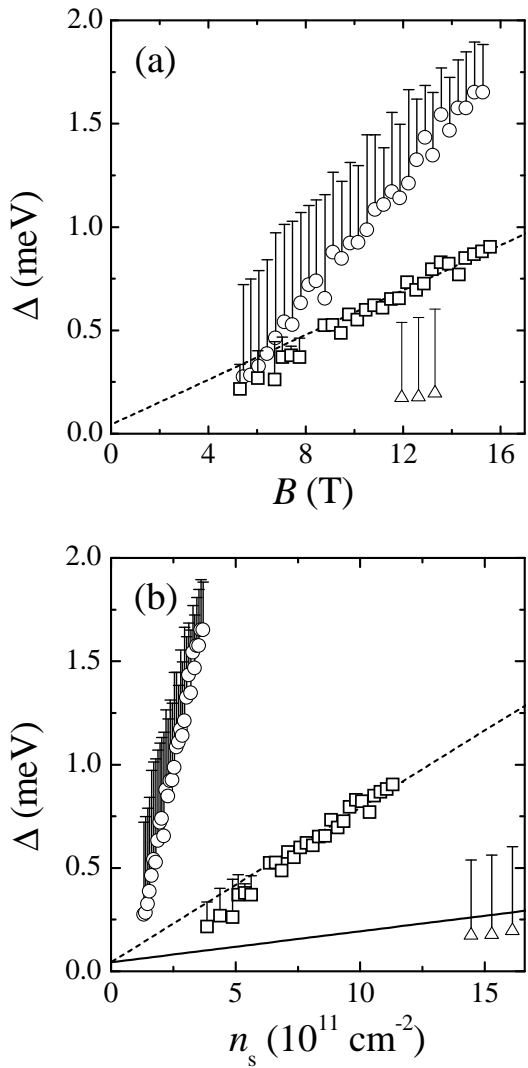}}
\caption{\label{fig3} The valley gap for $\nu=1$ (circles), $\nu=3$ 
(squares), and $\nu=5$ (triangles) as a function of magnetic field 
(a) and electron density (b). The systematic error bars correspond to 
the level width contribution. A linear fit of the $\nu=3$ data is 
shown by a dashed line. The solid line in (b) is the expected linear 
$n_s$ dependence of the single-particle valley gap \protect\cite{afs} 
for depletion layer charge density $1\times 
10^{11}$~cm$^{-2}$.}
\end{figure}

In Fig.~\ref{fig3}, we show the $\nu=1$ and $\nu=3$ valley splitting
measured in the low temperature limit as a function of both magnetic
field and electron density. For both filling factors, the value of
valley gap turns out to be strongly enhanced compared to the
single-particle value \cite{afs} (solid line in Fig.~\ref{fig3}(b)).
The valley gap enhancement decays rapidly with $\nu$ so that almost
no enhancement is observed for $\nu=5$. To our surprise, the data are
best described by a {\em linear} increase of the valley gap with $B$
(or $n_s$). This fact is reliably established for $\nu=3$ where the
geometric capacitance $C_0$ is attained in the entire range of
magnetic fields used except for the lowest $B$, which is indicated by
systematic error bars in Fig.~\ref{fig3} that correspond to the level
width contribution. Extrapolation to $B=0$ of the linear magnetic
field dependence of the $\nu=3$ gap yields a value which is
consistent with the single-particle splitting. For filling factor
$\nu=1$, the dependence of $\Delta$ on $B$ (or $n_s$) can also be
described by a linear function, although the experimental uncertainty
(estimated from data dispersion) of the data for $\nu=1$ is markedly
larger than that for $\nu=3$.

The linear magnetic field dependence of the enhanced valley gap
observed in silicon MOSFETs is similar to the puzzling proportional
$B$ dependence of the enhanced spin gap in the 2D electron system in
AlGaAs/GaAs heterostructures \cite{usher,aristov}. Before making
comparison with existing theories, we would like to emphasize that
they are expected to hold in the limit of weak electron-electron
interactions, i.e., the Coulomb energy, $E_c=e^2/\varepsilon l$
(where $l=(\hbar c/eB)^{1/2}$ is the magnetic length), in a 2D
electron system should be small compared to the cyclotron energy,
$\hbar\omega_c$. For real 2D electron systems, one determines that in
a magnetic field of 10~T, the Coulomb and the cyclotron energies are
approximately equal to each other for the 2D electrons in GaAs,
whereas in Si MOSFETs the energy $E_c$ exceeds $\hbar\omega_c$ by a
factor of 4. Therefore, even for the case of GaAs, the validity of
the Landau-level-based considerations
\cite{sondhi,yang,ando,macdonald} is questionable. Formally, these
cannot be applied to the case of Si MOSFETs. Bearing this in mind, we
will make qualitative comparison of our results with the
aforementioned theories below.

Theoretically, the basic idea is that the gap enhancement is expected
to be controlled by electron-electron interactions whose strength is
characterized by the Coulomb energy $E_c$
\cite{sondhi,yang,ando,macdonald}. Hence, as long as the enhancement
dominates the single-particle gap value, $\Delta_0$, the many-body
enhanced gap should essentially be proportional to $B^{1/2}$. This
law is in contradiction to the experiment. Not to mention that the
predicted values of the many-body gap are an order of magnitude
larger than the experimental ones.

Two approaches to introduce corrections to the $B^{1/2}$ law have
been formulated. Based on the traditional theory of exchange-enhanced
gaps \cite{ando}, the authors of Ref.~\cite{macdonald} took account
of the corrections to the exchange energy due to level overlap and
finite 2D layer thickness as well as the correlation energy
contribution. Knowing that in our case the level overlap is small
\cite{aristov}, we have verified that the approach \cite{macdonald}
is unable to noticeably increase the power of the theoretical
$B^{1/2}$ dependence of the gap.

In accordance with the other approach, in the limit $E_c\gg\Delta_0$,
a possible formation of the isospin textures --- skyrmions
\cite{sondhi} or merons \cite{yang} --- is expected to lead to a
reduction of the exchange-enhanced gap which is related to single
isospin-flip excitations. We have verified that in the range of
magnetic fields (or electron densities) used, the predicted crossover
between the two regimes as governed by the ratio $\Delta_0/E_c$
\cite{sondhi} has a weak effect on the theoretical square-root
dependence $\Delta(B)$. Although the actual symmetry of the isospin
system in Si MOSFETs is unknown, it is unlikely that the outcomes for
the two textures \cite{sondhi,yang} would be very different, because
the layer separation is small.

Thus, for our case both approaches fail to modify appreciably the
$B^{1/2}$ law and, therefore, they are not even able to give a
qualitative account of our experimental data.

In summary, we have performed low-temperature measurements of the
chemical potential jump across the valley gap at the lowest $\nu=1$
and $\nu=3$ in a 2D electron system in silicon. The valley splitting
is found to be strongly enhanced compared to the single-particle
value, decaying with filling factor. The data are best described by a
linear increase of the valley gap with magnetic field. This result is
similar to the proportional $B$ dependence of the enhanced spin gap
for the 2D electrons in GaAs \cite{usher,aristov}.

Recently, in the limit $E_c\gg\hbar\omega_c$, a remarkably different
behavior of the many-body gap to create a charge-carrying spin
texture excitation at integer $\nu$ has been predicted \cite{iordan}:
$\Delta_{\text{st}}=\nu|Q|\hbar\omega_c$, where $Q$ is the integer
topological charge. If the exchange effects play a similar role for
the valley splitting \cite{afs}, the predicted linear $B$ dependence
of the gap is consistent with our findings. Still, both the gap value
and its dependence on filling factor are not explained by the theory
\cite{iordan}.

We gratefully acknowledge discussions with S.~V. Iordanskii, A.
Kashuba, and S.~V. Kravchenko. This work was supported by RFBR grants
00-02-17294, 01-02-16424, and 01-02-26666, the Programmes
``Nanostructures'' under grant 97-1024 and ``Statistical Physics''
from the Russian Ministry of Sciences, and the Deutsche
Forschungsgemeinschaft under SFB grant 348.

%\begin{figure}
%\caption{\label{fig1}}
%\end{figure}

%\begin{figure}
%\caption{\label{fig2}}
%\end{figure}

%\begin{figure}
%\caption{\label{fig3}}
%\end{figure}

\end{document}